\documentclass[twocolumn,trackchanges]{aastex701}
\usepackage{hyperref}

\begin{document}
\title{XRISM Observations of The Prototypical Cold Front in Abell~3667}

\author[orcid=0009-0009-9196-4174,gname='Yuki',sname='Omiya']{Yuki Omiya}
\affiliation{Graduate School of Science, Nagoya University, Aichi 464-8602, Japan}
\email[show]{omiya\_y@u.phys.nagoya-u.ac.jp}  

\author[orcid=0000-0002-6102-1441,gname='Yuto',sname='Ichinohe']{Yuto Ichinohe}
\affiliation{RIKEN Nishina Center, Saitama 351-0198, Japan}
\email{ichinohe@ribf.riken.jp}

\author[orcid=0000-0003-2930-350X,gname='Kazuhiro',sname='Nakazawa']{Kazuhiro Nakazawa}
\affiliation{Kobayashi-Maskawa Institute for the Origin of Particles and the Universe (KMI), Furo-cho, Chikusa-ku, Nagoya, Aichi 464-8601, Japan}
\affiliation{Graduate School of Science, Nagoya University, Aichi 464-8602, Japan}
\email{nakazawa@u.phys.nagoya-u.ac.jp}

\author[orcid=0009-0001-3135-0754,gname='Hisamitsu',sname='Awaki']{Hisamitsu Awaki}
\affiliation{Department of Physics, Ehime University, Ehime 790-8577, Japan}
\email{awaki@astro.phys.sci.ehime-u.ac.jp}

\author[orcid=0000-0001-7917-3892,gname='Dominique',sname='Eckert']{Dominique Eckert}
\affiliation{Department of Astronomy, University of Geneva, Versoix CH-1290, Switzerland}
\email{Dominique.Eckert@unige.ch}

\author[orcid=0000-0003-0058-9719,gname='Yutaka',sname='Fujita']{Yutaka Fujita}
\affiliation{Department of Physics, Tokyo Metropolitan University, Tokyo 192-0397, Japan}
\email{y-fujita@tmu.ac.jp}

\author[orcid=0000-0003-3518-3049,gname='Isamu',sname='Hatsukade']{Isamu Hatsukade}
\affiliation{Faculty of Engineering, University of Miyazaki, 1-1 Gakuen-Kibanadai-Nishi, Miyazaki, Miyazaki 889-2192, Japan}
\email{hatukade@cs.miyazaki-u.ac.jp}

\author[orcid=0000-0003-0144-4052,gname='Maxim',sname='Markevitch']{Maxim Markevitch}
\affiliation{NASA / Goddard Space Flight Center, Greenbelt, MD 20771, USA}
\email{maxim.markevitch@nasa.gov}

\author[orcid=0000-0002-7031-4772,gname='Fran\c{c}ois',sname='Mernier']{Fran\c{c}ois Mernier}
\affiliation{IRAP, CNRS, Université de Toulouse, CNES, UT3-UPS, Toulouse, France}
\email{francois.mernier@irap.omp.eu}

\author[orcid=0000-0002-9901-233X,gname='Ikuyuki',sname='Mitsuishi']{Ikuyuki Mitsuishi}
\affiliation{Graduate School of Science, Nagoya University, Aichi 464-8602, Japan}
\email{mitsuisi@u.phys.nagoya-u.ac.jp}

\author[orcid=0000-0002-2784-3652,gname='Naomi',sname='Ota']{Naomi Ota}
\affiliation{Department of Physics, Nara Women’s University, Nara 630-8506, Japan}
\email{naomi@cc.nara-wu.ac.jp}

\author[orcid=0000-0002-9714-3862,gname='Aurora',sname='Simionescu']{Aurora Simionescu}
\affiliation{SRON Netherlands Institute for Space Research, Leiden, The Netherlands}
\email{a.simionescu@sron.nl}

\author[orcid=0000-0002-7962-4136,gname='Yuusuke',sname='Uchida']{Yuusuke Uchida}
\affiliation{Institute of Space and Astronautical Science (ISAS), Japan Aerospace Exploration Agency (JAXA), Kanagawa 252-5210, Japan}
\email{yuuchida@rs.tus.ac.jp}

\author[orcid=0000-0001-6252-7922,gname='Shutaro',sname='Ueda']{Shutaro Ueda}
\affiliation{Kanazawa University, Kanazawa, 920-1192 Japan}
\email{shutaro@se.kanazawa-u.ac.jp}

\author[orcid=0000-0001-7630-8085,gname='Irina',sname='Zhuravleva']{Irina Zhuravleva}
\affiliation{Department of Astronomy and Astrophysics, University of Chicago, Chicago, IL 60637, USA}
\email{zhuravleva@astro.uchicago.edu}

\author[orcid=0000-0003-3175-2347,gname='John',sname='Zuhone']{John Zuhone}
\affiliation{Center for Astrophysics | Harvard-Smithsonian, Cambridge, MA 02138, USA}
\email{john.zuhone@cfa.harvard.edu}


\begin{abstract}
We present high-resolution X-ray spectroscopy of the merging galaxy cluster Abell~3667 with \textit{XRISM}/Resolve. Two observations, targeting the cluster X-ray core and the prototypical cold front, were performed with exposures of 105~ks and 276~ks, respectively. We find that the gas in the core is blueshifted by $v_z\sim-200$~km~s$^{-1}$ relative to the brightest cluster galaxy, while the low-entropy gas inside the cold front is redshifted by $v_z\sim 200$~km~s$^{-1}$. 
As one moves further off-center across the front, the line-of-sight (LoS) velocity changes significantly, by $\Delta v_z=535^{+167}_{-154}$~km~s$^{-1}$, back to the value similar to that in the core. 
There are no significant LoS velocity gradients perpendicular to the cluster symmetry
axis.
These features suggest that the gas forming the cold front is flowing in the plane oriented along the LoS, supporting 
an offset merger scenario in which the main cluster has passed in front of the subcluster and induced rotation of the core gas in the plane perpendicular to the sky. \color{black}
The region just inside the front exhibits the largest LoS velocity dispersion seen across two pointings, $\sigma_z\sim420$~km~s$^{-1}$, which can be interpreted as a developing turbulence or a projection of the LoS velocity shear within the front. 
The large LoS velocity jump across the cold front, combined with the lack of Kelvin–Helmholtz instability on the surface of the front, suggests some mechanism to suppress it. 
For example, a magnetic field with $B>5\,\mu$G
is required if the cold front is stabilized by magnetic draping.
\end{abstract}


\section{Introduction}
Galaxy clusters are the most massive gravitationally bound systems.
Cluster mergers dissipate copious amounts of kinetic energy,
heating the intracluster medium (ICM), producing shocks and turbulence, accelerating cosmic ray particles and amplifying magnetic fields in the ICM \citep[e.g., ][]{2002ASSL..272....1S, 2014IJMPD..2330007B, 2019A&A...621A..40E, 2023PASJ...75...37O}. X-ray observations have long revealed subcluster stripping, merger shocks, and sharp contact discontinuities known as cold fronts \citep[e.g., ][]{1994Natur.372..439B, 1999ApJ...511...65J, 2002ApJ...567L..27M, 2007PhR...443....1M}, while radio observations show radio halos and relics generated by the intracluster cosmic rays and plausibly tracing merger-driven turbulence and shocks \citep[e.g., ][]{1998A&A...332..395E, 2010Sci...330..347V, 2012A&ARv..20...54F, 2014IJMPD..2330007B}. These multiwavelength signatures suggest vigorous gas motions. However, direct, precise spectroscopic measurements of the ICM velocities became possible only with nondispersive microcalorimeters, beginning with the \textit{Hitomi}/SXS \citep{2016SPIE.9905E..0VK,2016SPIE.9905E..0UT} 
%
observations of the Perseus cluster \citep{2016Natur.535..117H,2018PASJ...70....9H} 
and continuing with the \textit{XRISM}/Resolve spectrometer \citep{2025PASJ...77S...1T,2022SPIE12181E..1SI}. Resolve has already delivered results for several nearby clusters, including Abell~2029 \citep{2025ApJ...982L...5X}, Centaurus \citep{2025Natur.638..365X}, Coma \citep{2025ApJ...985L..20X}, Abell~2319 \citep{2025arXiv250805067X}, and Ophiuchus \citep{2025PASJ..tmp...89F}.

Abell~3667 (hereafter A3667) is a textbook example of a merging cluster with complex gas dynamics. It hosts one of the most prominent known cold fronts, extending over several hundred kiloparsecs with a sharp density and temperature discontinuity \citep{1999ApJ...521..526M,2001ApJ...551..160V,2024ApJ...973...98U}. The apparent lack of developed Kelvin–Helmholtz instability (KHI) at the front surface has been interpreted as evidence of suppressed transport due to magnetic draping and/or effective viscosity 
\citep{2001ApJ...549L..47V, 2011ApJ...743...16Z,2013MNRAS.436.1721R,2017MNRAS.467.3662I}. 

A3667 also exhibits luminous diffuse radio emission, including giant relics plausibly associated with merger shocks \citep{1997MNRAS.290..577R,2015MNRAS.447.1895R,2022A&A...659A.146D}. Optical and weak-lensing observations reveal multiple substructures
\citep{2009ApJ...704.1349O}. 

While early works interpreted the A3667 cold front as the front edge of a ram pressure-stripped merging subcluster that is moving along the symmetry axis approximately in the sky plane, an alternative possibility is that it is a giant sloshing front \citep{2007PhR...443....1M}, where the low-entropy gas beneath the front is spiraling around the core in the plane mostly perpendicular to the sky --- as seen in another projection in, e.g., A2319 \citep[][]{2021MNRAS.504.2800I}. Such sloshing could have been set off by an off-center merger. In a variation on this scenario, the spiraling cool gas could belong to the disturbing subcluster (rather than to the main cluster as in the sloshing scenario). Measuring the line-of-sight (LoS) gas velocity change across the front can unambiguously determine which merger geometry is correct, and we present such a measurement in this work.

Recently, \citet{2024A&A...689A.173O} analyzed deep \textit{XMM-Newton} data of A3667 and reported LoS velocity variations of a few hundred~km~s$^{-1}$ across the center, using the 
detector gain recalibration method of the line response 
proposed by \citet{2020A&A...633A..42S}.
These results provided the first quantitative evidence consistent with the offset merger scenario in this system. However, the constraints were limited by systematic uncertainties and by the inability of CCD spectra to 
measure turbulent broadening.
The advent of \textit{XRISM}/Resolve overcomes these limitations, enabling precise measurements of gas bulk motions and velocity dispersions in A3667.

In this paper, we report the A3667 results from \textit{XRISM}/Resolve. Throughout this paper,
we adopt a flat $\Lambda$CDM cosmology with $H_0=70~\mathrm{km~s^{-1}~Mpc^{-1}}$, $\Omega_{\mathrm{m}}=0.3$, and $\Omega_\Lambda=0.7$. At $z=0.055$, $1\arcmin$ 
corresponds to 64~kpc.
Unless otherwise noted, uncertainties are quoted at the $1\sigma$ level.

\section{Data treatment}

\subsection{Observations}

\begin{figure*}[ht!]
  \centering
  \begin{minipage}{1.0\columnwidth}
    \centering
    \includegraphics[width=\columnwidth]{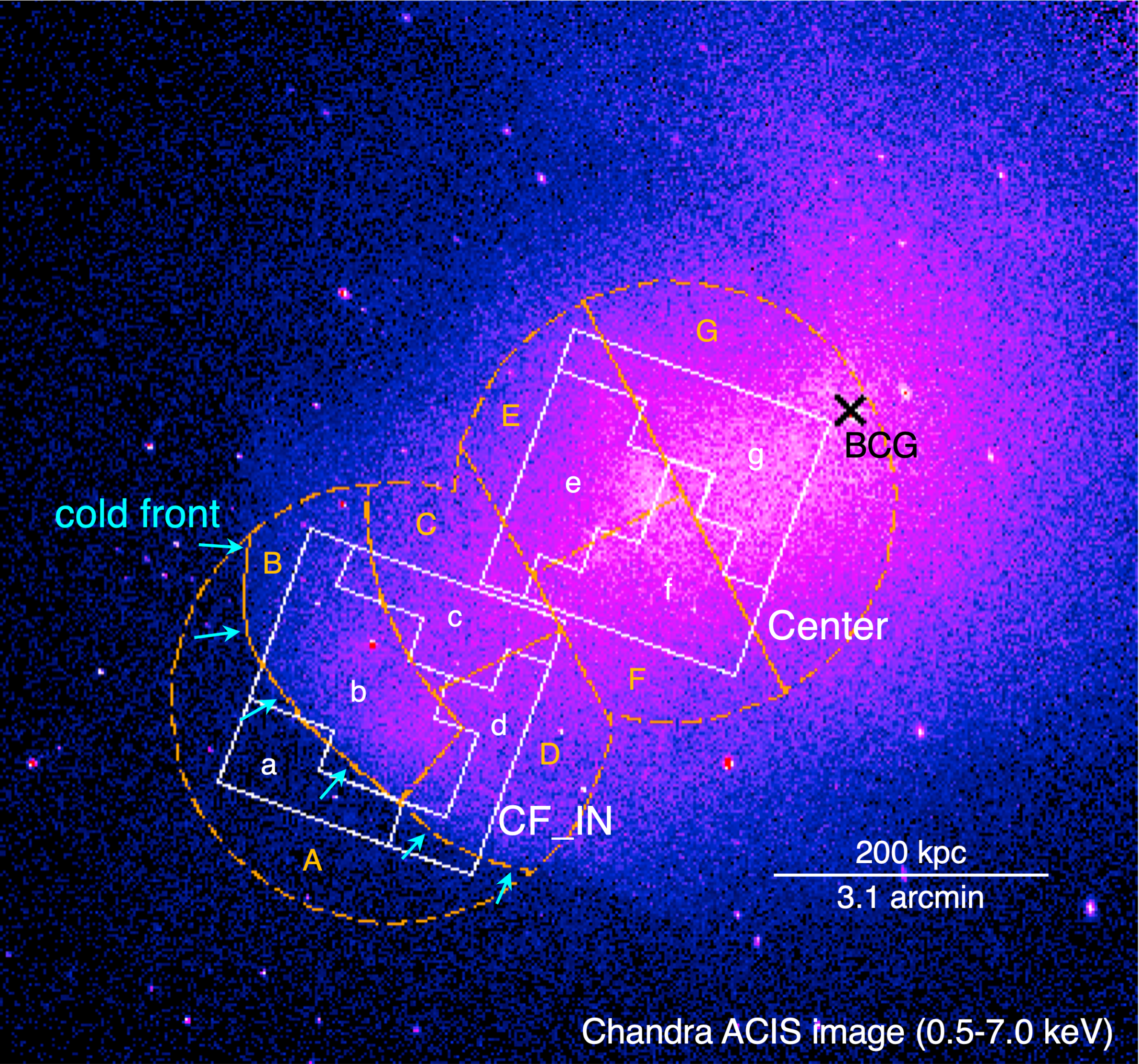}
  \end{minipage}
  \begin{minipage}{1.0\columnwidth}
    \centering
    \includegraphics[width=\columnwidth]{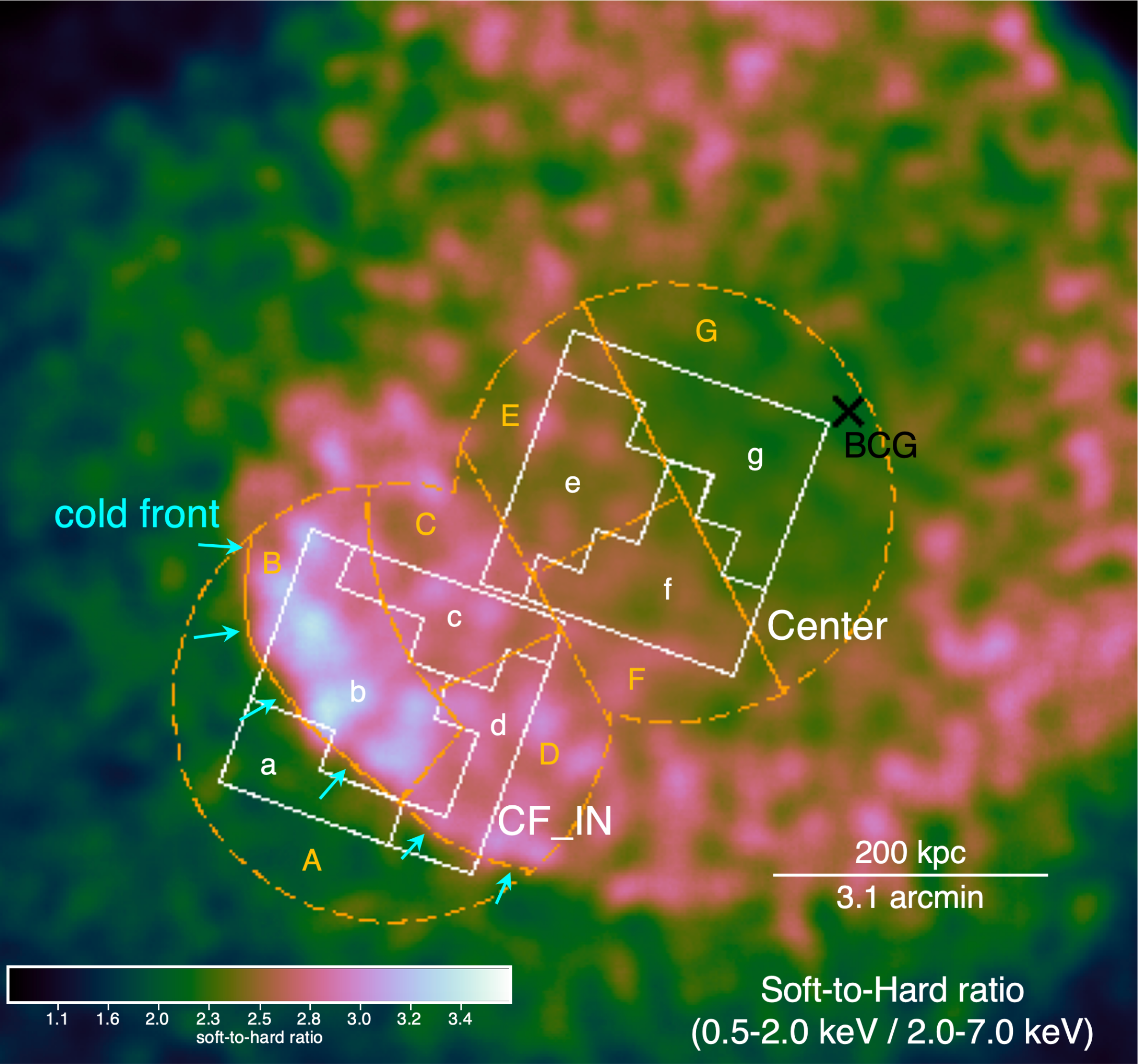}
  \end{minipage}
\caption{(Top) \textit{Chandra} image of A3667 in the $0.5-7.0$~keV band. White squares show fields of view of the two \textit{XRISM}/Resolve observations. White pixellated
subregions
\color{black}
within the white squares represent detector regions used for spectral extraction.
The dashed orange regions represent the sky regions for which the ICM parameters are derived. A black ``x'' marks the position of the BCG, and cyan arrows indicate the prominent cold front seen in the \textit{Chandra} image.
(Bottom) Soft-to-hard ratio map of A3667, constructed from the ratio of the 0.5–2.0 keV to $2.0–7.0$ keV \textit{Chandra} images.
This map shows spatial variations of the ICM temperature, with brighter regions corresponding to cooler gas.
{Alt text: The top panel shows an X-ray image of Abell 3667, with right ascension on the horizontal axis and declination on the vertical axis. The bottom panel shows a hardness-ratio map, where brighter areas correspond to cooler gas.
\label{fig:chandra_image}}}
\end{figure*}

\textit{XRISM}/Resolve conducted two observations of A3667 from 30 October to 4 November 2024. The Resolve field of view (FoV) is a $3'.1\times3'.1$ square covered by an array of $6\times 6$ pixels, each one returning a spectrum with an average resolution of 4.5 eV (FWHM), \citep{2025PASJ...77S...1T}.
The two observed fields are indicated by white boxes in Figure~\ref{fig:chandra_image}. One pointing covered the edge of the cold front (\texttt{CF\_IN}; OBSID: 201050010, RA, Dec = 303.26525, -56.88616; PI: Ichinohe), and the other targeted the cluster center (\texttt{Center}; OBSIDs: 201051010, RA, Dec = 303.17404, -56.84815; PI: Ichinohe).

\begin{deluxetable*}{lcccc}
\tablecaption{Best-fit Parameters of Full Array in 2.0--10.0~keV and 6.2--6.7~keV Energy Ranges \label{tab:FoV_parameter}}
\tablehead{
\colhead{Parameter} & \multicolumn{2}{c}{CF\_IN} & \multicolumn{2}{c}{Center} \\
\colhead{} & \colhead{2.0--10.0 keV} & \colhead{6.2--6.7 keV} & \colhead{2.0--10.0 keV} & \colhead{6.2--6.7 keV}
}
\startdata
$kT$ (keV) & $4.89^{+0.12}_{-0.12}$ & $4.89^{+0.29}_{-0.28}$ & $5.94^{+0.17}_{-0.16}$ & $5.71^{+0.32}_{-0.31}$ \\
Abundance ($Z_{\odot}$) & $0.45^{+0.02}_{-0.02}$ & $0.41^{+0.02}_{-0.02}$ & $0.40^{+0.03}_{-0.02}$ & $0.32^{+0.04}_{-0.03}$ \\
Redshift ($\times 10^{-2}$) & $5.6048^{+0.0092}_{-0.0083}$ & \nodata & $5.4970^{+0.0063}_{-0.0056}$ & \nodata \\
Relative Velocity (km~s$^{-1}$)\tablenotemark{a} & $+125^{+28}_{-25}$ & \nodata & $-198^{+19}_{-17}$ & \nodata \\
Velocity Dispersion (km~s$^{-1}$) & $292^{+26}_{-24}$ & \nodata & $180^{+17}_{-16}$ & \nodata \\
C-stat/d.o.f. & $15970/15994$ & $1071/994$ & $15322/15994$ & $1089/994$ \\
\enddata
\tablenotetext{a}{Velocity relative to BCG.}
\end{deluxetable*}

\subsection{Data reduction and Spectral modeling}
\label{Data reduction and Spectral modeling}

Data reduction was performed following the procedure in \citet{2025arXiv251016553O} 
using XRISM CALDB version 20250315. 
\color{black}
We processed the data with \texttt{xapipeline} in HEAsoft v6.34 (CALDB v11), applied the screening criteria, and generated RMFs with \texttt{rslmkrmf} and ARFs with \texttt{xaarfgen} using the Chandra image in the $2.0–7.0$~keV band.
After screening, the net exposures were 105~ks for \texttt{Center} and 276~ks for \texttt{CF\_IN}. Instrumental systematics were estimated using the in-orbit $^{55}$Fe calibration data,
yielding uncertainties of $\sim$15~km~s$^{-1}$ for line positions and $<2$~km~s$^{-1}$ for line broadening at the Fe K$\alpha$ line energies, both negligible relative to the statistical errors.

Spectral fitting was performed in \texttt{XSPEC} v12.15.0 \citep{1996ASPC..101...17A} with AtomDB~3.0.9, adopting protosolar abundances from \citet{2009LanB...4B..712L}. The ICM emission was modeled with \texttt{bapec} under collisional ionization equilibrium, with free parameters of temperature, metal abundance, redshift, and Gaussian velocity dispersion. Average heliocentric velocity corrections of $-22$~km~s$^{-1}$, derived using \texttt{barycen}, were applied to the fits for both pointings. Galactic absorption was modeled with \texttt{tbabs} with a fixed column density of $N_{\rm H}=1.4\times10^{21}$~cm$^{-2}$ \citep{2005A&A...440..775K}. Non–X-ray background (NXB) spectra were generated with \texttt{rslnxbgen} from night-Earth events using the Resolve NXB Database (v2).
They were
\color{black}
modeled with a power-law and Gaussian lines, and the best-fit model was included as the NXB component in the fits,
following the recommended procedures described in the ``Resolve NXB Database and Spectral Extraction Recipes'' on the HEASARC website.
\color{black}

\section{Results}

\subsection{Spectra from large regions}
\label{sec:FoV_spectra_of_Resolve}

\begin{figure*}[htpb]
\centering
\includegraphics[width=18cm]{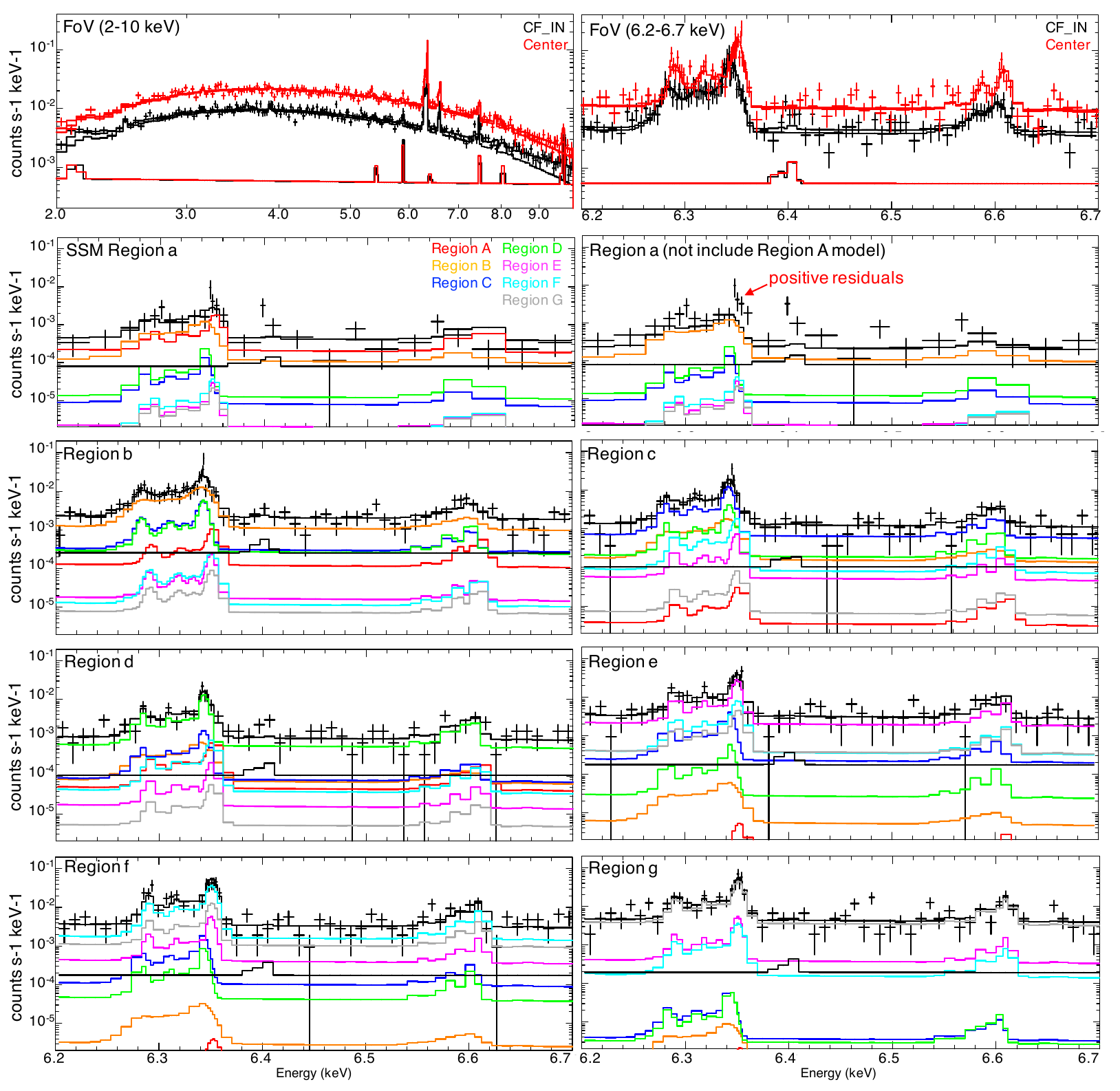}
\caption{(Top left) Resolve spectra in the 2.0--10.0~keV energy band for the cold front (\texttt{CF\_IN}, black) and central (\texttt{Center}, red) pointings. The best-fit models are overlaid on each spectrum together with NXB models
(Top right) Zoom-in of the Resolve spectra in the 6.2--6.7~keV energy band. Data are binned by 
2~eV for display purposes. 
(Bottom) The spectra in region~a--g used for the SSM analysis in the 6.2--6.7~keV energy band. Each colored curve shows the modeled emission from the corresponding sky region: region~A (red), B (orange), C (blue), D (green), E (magenta), F (cyan), and G (light gray).
The spectrum of region~a is also shown with the sum of all models (black) except for the region~A model, highlighting the blueshifted contribution of region~A relative to the others.
\color{black}
{Alt text: Ten line graphs. In the upper left panel, the x axis shows the energy from 2.0 to 10.0 kilo electron volt. The y axis shows the count from 0.0002 to 0.3 counts per second and per kilo electron volt. In the upper right panel, the x axis shows the energy from 6.2 to 6.7 kilo electron volt. The y axis shows the count from 0.0002 to 0.3 counts per second and per kilo electron volt.
In the bottom panels, the x axis shows the energy from 6.2 to 6.7 kilo electron volt. The y axis shows the count from 0.000002 to 0.2 counts per second and per kilo electron volt.
}}\label{fig:Resolve_spectra}
\end{figure*}

We first extracted spectra for the two full $3'.1$ 
Resolve
fields for the \texttt{CF\_IN} and \texttt{Center} pointings and fit them with single-temperature models in the $2.0–10.0$~keV band. As shown in Figure~\ref{fig:Resolve_spectra}, the models reproduce the spectra well.
To verify this result, we restricted the fits to the narrow $6.2-6.7$~keV interval covering the Fe K$\alpha$ 
lines of He- and H-like ions;
the derived temperatures were consistent with those from the broad-band fits within the statistical errors, as summarized in Table~\ref{tab:FoV_parameter}.

Relative to the optical redshift of the BCG  \citep[$z=0.05567$; ][]{2004AJ....128.1558S}, the LoS bulk velocities are measured as $+125^{+28}_{-25}$~km~s$^{-1}$ in \texttt{CF\_IN} and $-198^{+19}_{-17}$~km~s$^{-1}$ in \texttt{Center}, corresponding to a velocity difference of $323^{+33}_{-31}$~km~s$^{-1}$.
The velocity dispersions are $292^{+26}_{-24}$~km~s$^{-1}$ and $180^{+17}_{-16}$~km~s$^{-1}$, respectively.
The values are consistent with the range 
($170-300$~km~s$^{-1}$) 
inferred from the analysis of surface brightness fluctuations within the $\sim$400~kpc region of A3667 \citep{2024MNRAS.528.7274H}.
\color{black}

\subsection{Resolving the cluster velocity structure}
\label{Spatially resolved spectroscopy}

To investigate the spatial structure of gas motions while accounting for the finite angular resolution of the \textit{XRISM} mirror (whose point-spread function, PSF, has a half-power diameter of $\sim$1\farcm3), we conducted a spatial–spectral mixing (SSM) analysis following the multi-source ARF technique of \citet{2018PASJ...70....9H}. The input sky domain for ARF generation was restricted to a circular region with radius 2\farcm5 centered at each pointing position. This procedure ensures that photons scattered into the detector from adjacent bright regions are properly modeled.

The sky regions were chosen to probe interesting physical features in the cluster. Figure~\ref{fig:chandra_image} shows a \textit{Chandra} broad-band image and a 
soft-to-hard ratio (SHR) map (a ratio of
the 0.5--2.0~keV and 2.0--7.0 keV images), which shows variations of the projected ICM temperature.
The sky regions are indicated by the orange dashed contours in Figure~\ref{fig:chandra_image}. 
The region outside (on the less-dense side of) the cold front is defined
as region~A.
Immediately inside the front, we identify a 
bright cool region,
designated as region~B. Moving further northwest from the cold front along the cluser long axis, the SHR map reveals two step-like temperature changes (also seen in the temperature map in \citealt{2017MNRAS.467.3662I}). 
Regions C–G were defined to trace the successive steps 
in the SHR map, 
from the immediate vicinity of the cold front to the downstream outskirts, as well as to probe the thermodynamic and kinematic variations along the cluster’s long (merger) axis. Specifically, Regions~C and D lie just inside the cold front and are separated into the right (north) and left (south) sides relative to the inferred subcluster trajectory. Regions~E and F are located at intermediate distances and are similarly divided into two sides, while Region~G covers the outermost downstream sector.
\color{black}

On the detector plane, the two fields of view were divided into 
subregions so that the detector pixels closely match each sky region
\color{black}
(white jagged regions a--g in Figure~\ref{fig:chandra_image}). The region sizes are comparable to or larger than the PSF width. This ensures that each extracted spectrum is dominated by the emission from its associated sky region, while still probing gradients across the FoV. For each combination of sky region and detector segment, we generated cross-region ARFs using \texttt{xaarfgen}, weighting by the 2.0–7.0~keV \textit{Chandra} surface-brightness map and folding with the appropriate attitude solution. In the joint fitting procedure, all detector spectra were fit simultaneously: parameters associated with a given sky region 
(temperature, abundance, redshift, velocity dispersion, and normalization of the \texttt{bapec} model) 
were linked across detector spectra, enforcing a single physically consistent solution for that region.

The spectrum of the detector region a, which covers part of the sky region A, suffers from limited photon statistics and a significant scattered contribution from the adjacent, brighter region~B. The temperature determination based solely on the continuum is susceptible to systematic error. We therefore fixed the temperature of region~A at 8.0~keV and the abundance at 0.41~$Z_{\odot}$, 
based on results of \citet{2024A&A...689A.173O}.
As discussed in Appendix~\ref{sec:ssm_free_fitting}, this  does not alter our main conclusions.
\color{black}

\begin{figure*}[h]
  
  \centering
  \begin{minipage}{1.0\columnwidth}
    \centering
    \includegraphics[width=\columnwidth]{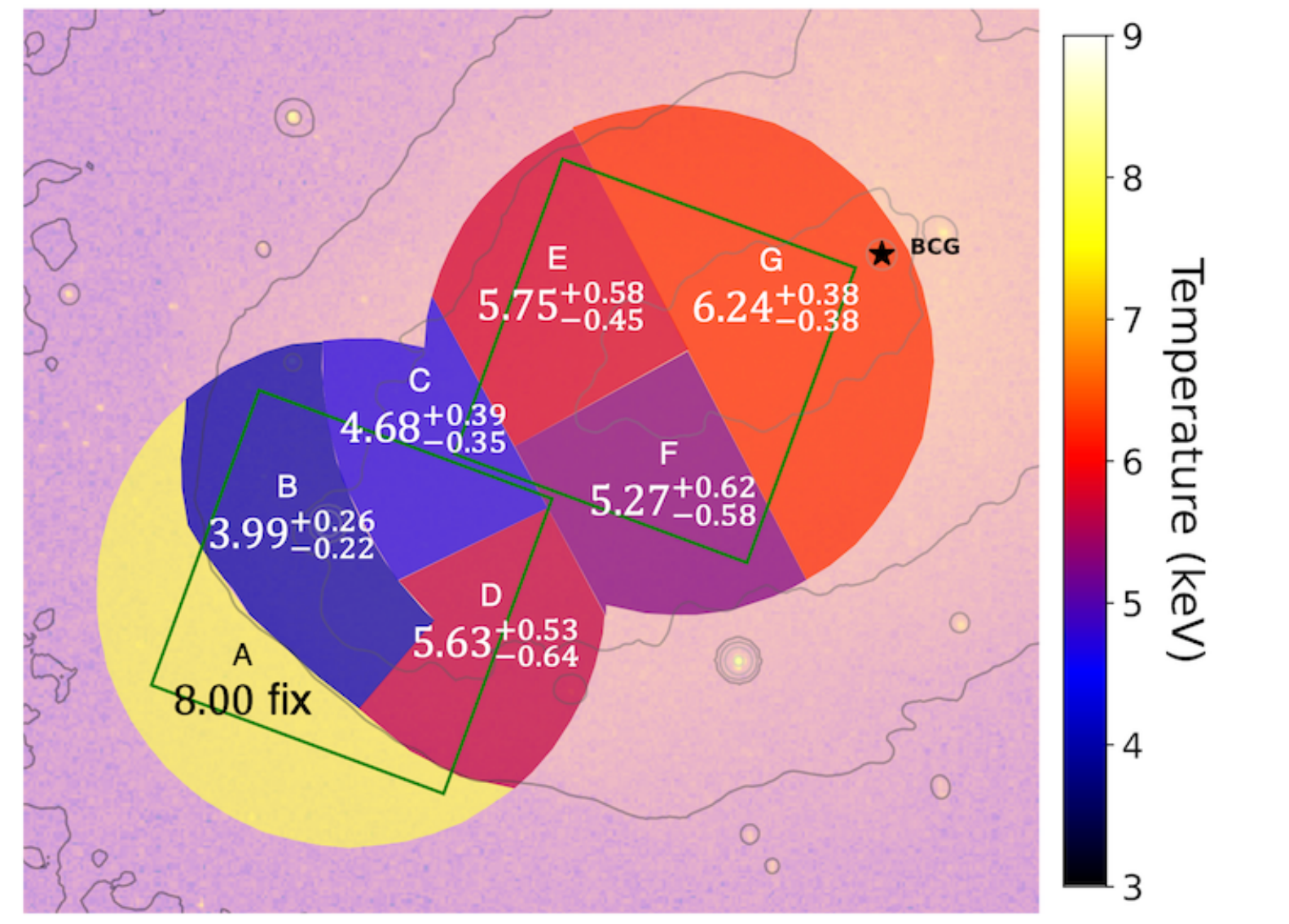}
  \end{minipage}
  \begin{minipage}{1.0\columnwidth}
    \centering
    \includegraphics[width=\columnwidth]{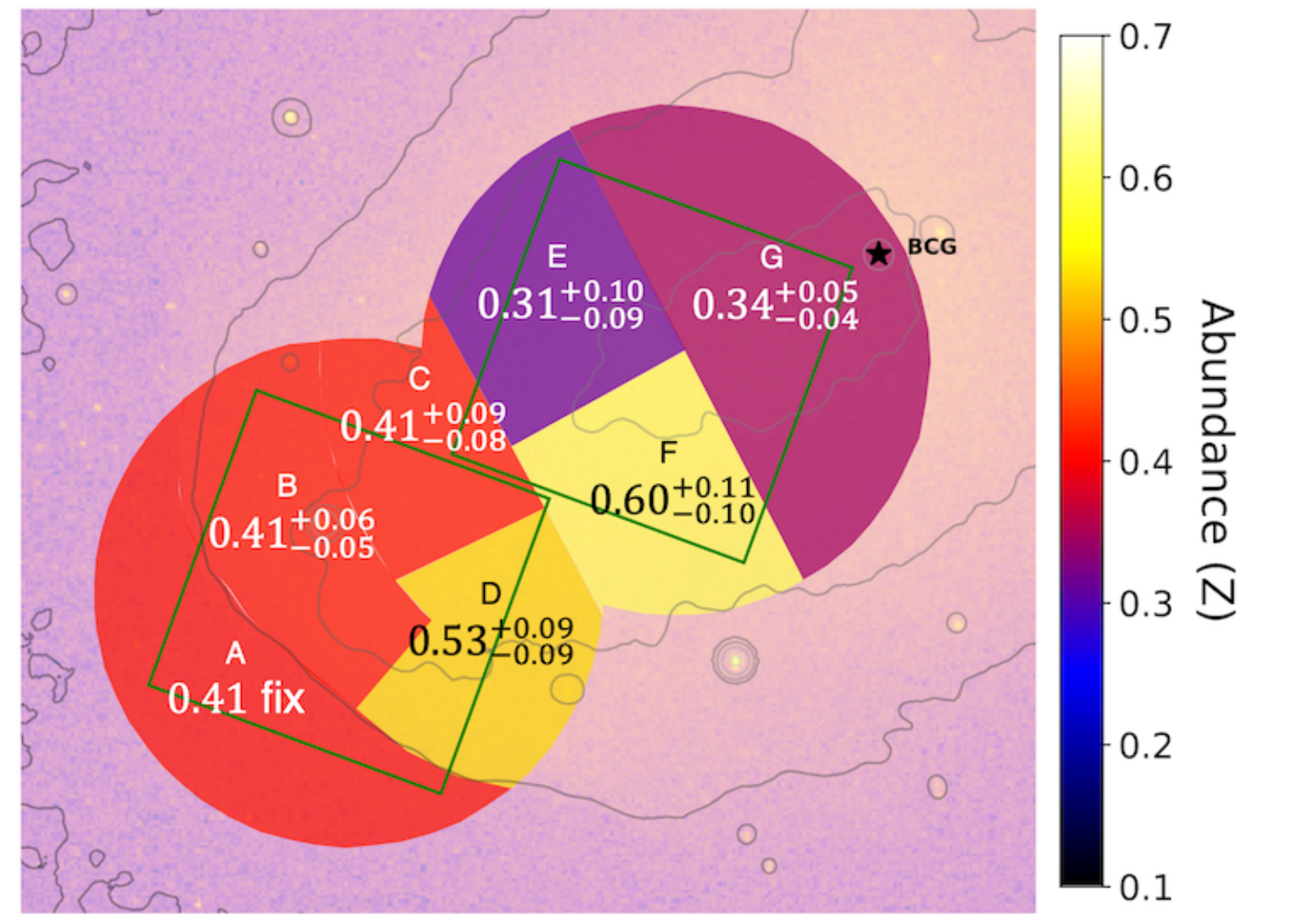}
  \end{minipage}
  
  \centering
  \begin{minipage}{1.0\columnwidth}
    \centering
    \includegraphics[width=\columnwidth]{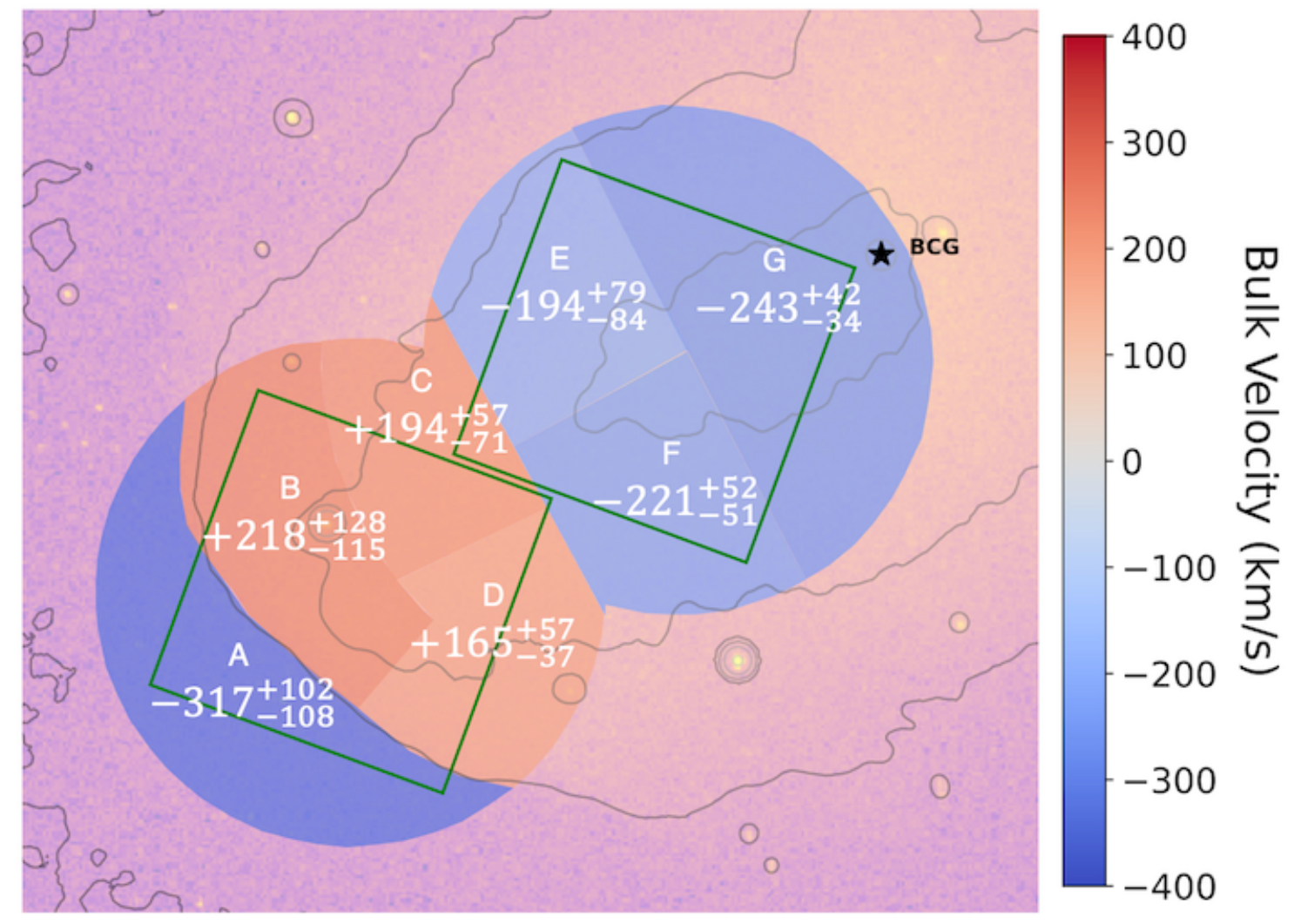}
  \end{minipage}
  \begin{minipage}{1.0\columnwidth}
    \centering
    \includegraphics[width=\columnwidth]{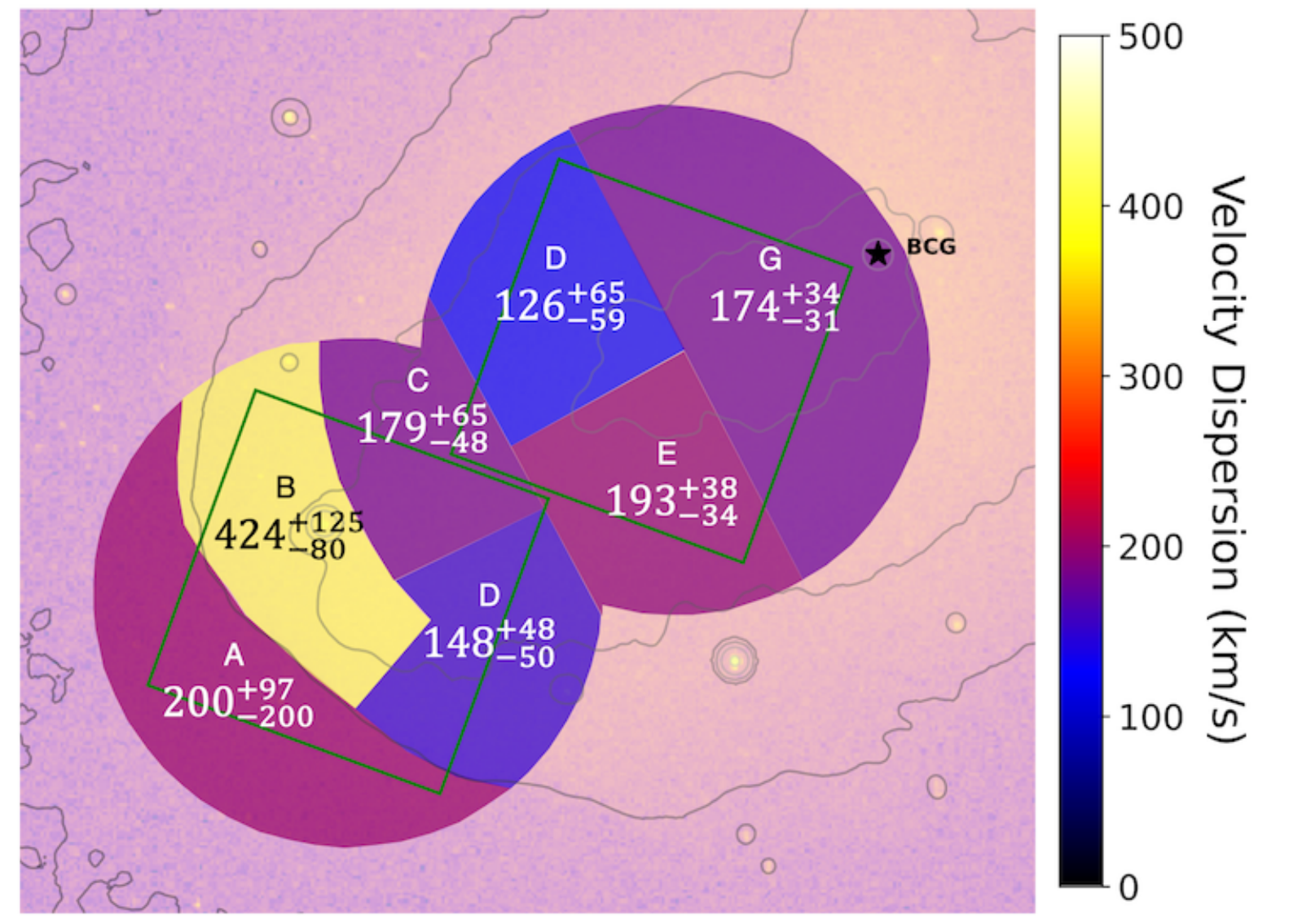}
  \end{minipage}
\caption{Maps of temperature (top left), abundance (top right), LoS bulk velocity relative to the BCG (bottom left), and LoS velocity dispersion (bottom right) for the sky regions defined in Figure~\ref{fig:chandra_image}. The black star indicates the position of the BCG. 
Contours represent the X-ray surface brightness
from the \textit{Chandra} image. The green boxes indicate regions covered by the \textit{XRISM}/Resolve observations.
{Alt text: Four color maps arranged in two rows and two columns, each with right ascension on the horizontal axis and declination on the vertical axis. The upper left map shows temperature in kilo–electron volts ranging from 3.0 to 9.0. The upper right map shows abundance in solar units ranging from 0.1 to 0.7. The lower left map shows bulk velocity in kilometers per second ranging from minus 400 to plus 400. The lower right map shows velocity dispersion in kilometers per second ranging from 0 to 500.}}\label{fig:Resolve_map_ssm}
\end{figure*}

The resulting maps of temperature, abundance, redshift, and velocity dispersion are shown in Figure~\ref{fig:Resolve_map_ssm}, with the corresponding spectra and models shown in Figure~\ref{fig:Resolve_spectra}. Our temperatures follow the \textit{Chandra}-based SHR map: region~B, just inside the cold front, is the coolest, while the downstream
region~G
is the hottest,
consistent with the trends reported from \textit{Chandra} data.


The LoS bulk velocities reveal a coherent pattern across the cold front. The region just inside the front (region~B) is redshifted 
by $v_z=218^{+128}_{-115}$~km~s$^{-1}$ w.r.t.\ the BCG, 
regions~C and D show a similar redshift, while regions E–G closer to the center are blueshifted by 
$v_z\approx -200$~km~s$^{-1}$ 
relative to the BCG. The region outside the front (region~A) has $v_z=-317^{+102}_{-108}$~km~s$^{-1}$, similar to the core gas velocities (E-G). Across the cold front interface, the velocity difference between regions~A and B reaches $535^{+167}_{-154}$~km~s$^{-1}$, and the pattern is such that the gas under the front (regions B, C, and D) is moving along the LoS relative to the rest of the cluster.
Although the exact regions differ, the velocity in region A outside the front is consistent with the southeast sector identified by \citet{2024A&A...689A.173O}, where the \textit{XMM} data for the exterior likewise suggested a net blueshift.

The velocity dispersions are broadly uniform at the $\sim$150–200~km~s$^{-1}$ level across most regions, consistent within their 1$\sigma$ confidence intervals. An exception is region~B with a significantly higher dispersion, $\sigma_z=424^{+125}_{-80}$~km~s$^{-1}$. 
This value is comparable to the high dispersion reported in A2319, another merger with a prominent cold front \citep{2025arXiv250805067X}.

\section{Discussion}
\subsection{Velocities and kinetic pressure}
\label{sec:vel_kinetic_pressure}

The Resolve fits presented above reveal a clear LoS velocity change across the cold front --- from $v_z\approx -300$~km~s$^{-1}$ outside the front to $v_z\approx +200$~km~s$^{-1}$ inside.
To place this velocity shear in a dynamical context, we convert it to a Mach number using the sound speed $c_s = (\gamma k_{\rm B}T/\mu m_p)^{1/2}$ with the adiabatic index $\gamma=5/3$ and the mean molecular weight $\mu=0.60$. The A--B shear corresponds to a Mach number 
$\mathcal{M} = 0.37^{+0.10}_{-0.10}$, calculated using the temperature of region A.

Our LoS velocity dispersions are broadly uniform at $\sigma_z\sim150-200$~km~s$^{-1}$, except for region~B under the front, which shows $\sigma_z\approx 400$~km~s$^{-1}$.
Assuming isotropic turbulence, the LoS dispersion corresponds to a 3D Mach number $\mathcal{M}_{\rm 3D} = \sqrt{3}\,\sigma_z / c_s$.
The associated nonthermal pressure fraction due to turbulence is given by \citep[e.g.,][]{2019A&A...621A..40E}
\begin{equation}
\frac{P_{\rm NT}}{P_{\rm tot}}
=\frac{\mathcal{M}_{\rm 3D}^2}{\mathcal{M}_{\rm 3D}^2+3/\gamma}.
\end{equation}
Most of the cluster regions show $P_{\rm NT}/P_{\rm tot} \simeq 1.7$--$4.5$\%, while region~B reaches a significantly higher value of 
$22\pm8$\%.
This 
indicates that region B is more dynamically disturbed, consistent with the sloshing or bulk-flow motions discussed in Section~\ref{sec:Implications_for_the_offset_merger_scenario}.
\color{black}


\subsection{Galaxy–ICM velocity comparison}
\label{sec:Galaxy--ICM_Velocity_Comparison}

The gas dynamics observed with \textit{XRISM}/Resolve reveal coherent LoS bulk flows and the localized
high velocity dispersion
across the cold front, indicating complex ICM motion during the merger. An important question is whether these ICM motions are coupled with galaxies.
\citet{2009ApJ...693..901O} applied the KMM (Kaye's Mixture Model) analysis to identify multiple galaxy substructures.
Three subgroups of particular interest are KMM2, KMM4, and KMM5
\cite[see Figure~11 in ][]{2009ApJ...693..901O}.
KMM2 lies in the northwest, centered on the second-brightest cluster galaxy (2nd BCG). 
It exhibits a redshifted average peculiar velocity of $+362$~km~s$^{-1}$. KMM5 is centered on the BCG itself and is distributed across the south-eastern part of the cluster, with a blueshifted average peculiar velocity of $-163$~km~s$^{-1}$. KMM4 lies just to the south-east of the cold front and shows a stronger blueshift of $-698$~km~s$^{-1}$. 

KMM5 matches the ICM velocity pattern measured in regions~E-G, which lie in the same south-eastern direction and exhibit blueshifts of approximately $-200$~km~s$^{-1}$. The close spatial and kinematic agreement suggests that the south-eastern galaxy subgroup and its surrounding ICM are moving together as a dynamically coupled structure. Such co-movement indicates that this subcomponent of the cluster may be infalling as a coherent unit during the ongoing merger, maintaining both gravitational and hydrodynamic coherence.
On the other hand, regions~B-D are redshifted and are located just inside the cold front. These regions appear to be kinematically decoupled from the surrounding 
galaxies.
This behavior is consistent with the sloshing-core (offset-merger) scenario, in which the gas in these decoupled regions follows a distinct velocity field driven by sloshing motions, rather than being directly coupled to the surrounding galaxy population, as discussed in Section~\ref{sec:Implications_for_the_offset_merger_scenario}.

The KMM4 subgroup is located just outside the cold front in the south-eastern region and spatially encompasses ICM region~A. Both KMM4 and region~A exhibit significant blueshifts, $-698$~km~s$^{-1}$ and $-296^{+102}_{-108}$~km~s$^{-1}$ respectively, indicating that they are moving in the same direction along the LoS. This agreement in sign suggests that the galaxy subgroup and surrounding gas may share a common bulk flow component.

\subsection{Implications for the offset merger scenario}
\label{sec:Implications_for_the_offset_merger_scenario}

The LoS velocity field revealed by \textit{XRISM}/Resolve provides clear evidence supporting the previously proposed offset-merger scenario for A3667 \citep{2024A&A...689A.173O,2016arXiv160607433S,1999ApJ...518..603R}. 
A3667 has long been interpreted as a head-on merger, due to the small LoS velocity difference between the BCG and the 2nd BCG ($\sim$120~km~s$^{-1}$; \citealt{2009ApJ...704.1349O}), and its sharp X-ray cold front and symmetric radio relics \citep{2001ApJ...551..160V}. 
A head-on scenario would yield spatially and kinematically symmetric patterns with modest LoS velocity gradients. 
In contrast, the 
Resolve
data reveal a LoS velocity change of nearly 600~km~s$^{-1}$ across the cold front and another change by $\sim$400~km~s$^{-1}$ between the sloshing core and the central ICM. 
This velocity pattern argues instead for a merger with a significant impact parameter that imparts angular momentum to the core and drives sloshing and large-scale flows in the plane perpendicular to the sky.

\begin{figure}[ht!]
\plotone{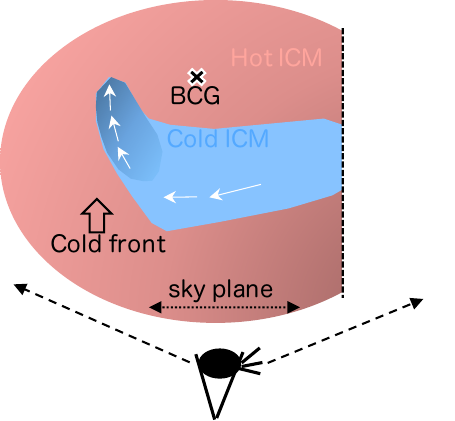}
\caption{A schematic illustration of the gas flows in A3667. The cool, low-entropy gas associated with the displaced main core (or, alternatively, the core of the disturber subcluster) is shown in blue, while the hot, high-entropy ICM of the main cluster is shown in red. White arrows indicate the inferred directions of the bulk gas motion, including the rotation of the cool gas.
The viewing direction is from the bottom of the figure.
{Alt text: A color schematic of the merging scenario in Abell~3667.
\label{fig:schematic}}}
\end{figure}

Offset merger simulations predict the generation of large-scale bulk flows, rotational motions, and cold fronts \citep[e.g.,][]{2006ApJ...650..102A,2011MNRAS.413.2057R}, all of which are consistent with our observations. In particular, the low-entropy gas spanning regions~B–D shows motions distinct from the surrounding ICM, naturally interpreted as a sloshing core that has acquired angular momentum during the encounter. The viewing geometry is illustrated in Figure~\ref{fig:schematic}. 
The cool gas forming the cold front can (1) originate in the displaced core of the main cluster centered on the BCG, or (2) it can be the core of the disturber subcluster. Scenario 1 is the analog of A2319 \citep[e.g., ][]{2021MNRAS.504.2800I}, whose core is seen sloshing in the sky plane. Looking at Figure~1 of \citet{2021MNRAS.504.2800I}, we note that if the observer were looking at the A2319 core from the side, near the skyplane (which corresponds to the A3667 orientation), the sharpness of the cold front in the X-ray image would be maintained for a large range of viewing angles. In this scenario, the disturbing subcluster could be the ``mushroom'' seen far to the northwest in the XMM-Newton image of A3667 \citep{2016arXiv160607433S} toward the 2nd BCG. In scenario 2, where region B is the disturber subcluster gas again seen from an orthogonal direction, 
the solid angle subtended by the sharp cold front at the head of the subcluster would be rather small (see, e.g., the nose of the bullet in the Bullet Cluster, \citealt{Markevitch06}), so a fine-tuned viewing geometry would be needed to produce the observed very sharp front in A3667.


The observed small velocity difference between the BCG and the 2nd BCG 
can be explained if they are presently located near the apocenters of their post-passage orbits \citep{2024A&A...689A.173O}. Thus, the velocity field and the galaxy dynamics can be naturally reconciled within the offset–merger framework.

Further insight into the sloshing core’s dynamics comes from the geometry of its rotation. Regions E–F (north and south of the core) and C–D show no significant LoS velocity differences. If the rotation axis were strongly inclined with respect to the LoS, we would expect to observe opposite-sign velocity shifts (i.e., a shear) between these regions due to the projection of tangential motion.
The absence of such a difference indicates an almost edge-on view of the sloshing motion (where sloshing occurs in the plane perpendicular to the sky).
Hydrodynamical simulations of offset mergers predict rotational velocities of $\sim$300–400~km~s$^{-1}$ within the sloshing core \citep[e.g.,][]{2010ApJ...717..908Z}, similar to the $\sim$400~km~s$^{-1}$ velocity difference between the sloshing core (B–D) and central ICM (E–G).

Region~B also exhibits the largest velocity dispersion in the field, $\sigma_z\approx400$~km~s$^{-1}$. This enhanced broadening may not arise from fully developed turbulence alone. Local velocity shear in the cool gas under the cold front, associated with rotational motion of the sloshing core, could trigger Kelvin–Helmholtz instabilities (KHI), generating small-scale turbulence that broadens the velocity distribution. The dispersion may also reflect the superposition of different LoS velocities constituting the same shear, 
or other unresolved large-scale merger-driven flows. Thus, the observed broadening likely represents a mixture of turbulent and bulk kinematic contributions
\citep[e.g.,][]{2016ApJ...821....6Z}. 

\subsection{Constraints on magnetic field strength}
\label{Constraints_on_magnetic_field_strength}

The cold front in A3667 is remarkably sharp.
This is noteworthy given the presence of a substantial velocity shear across the front (i.e., the layers of gas moving tangentially with respect to each other), as directly measured by the 
Resolve
observations, and as inferred earlier from the X-ray image \citep{2001ApJ...551..160V}.
At such a layer boundary, any small perturbation should quickly grow because of KHI. If the growth time is shorter than the time it takes for the perturbation to move away from the visible sector of the front, the instability will broaden the interface (e.g., \citealt{2002AstL...28..495V, 2004MNRAS.350L..52C}). Chandra imaging has indeed revealed small-amplitude perturbations of the front shape on a $\lambda\sim 100$ kpc scale in the plane of the sky and possibly along the LoS \citep{2017MNRAS.467.3662I}, but those perturbations have not grown to an amplitude sufficient to destroy the sharp interface. 

The growth of KHI at small linear scales should thus be suppressed by some physical mechanism. Several have been proposed for this front, including diffusion, magnetic field tension, viscosity, and gravity.
The gravitational acceleration across the front, estimated by \citet{2002AstL...28..495V}, may help suppress Rayleigh–Taylor and KH instabilities by providing a restoring force against the deceleration of the dense cold gas, but does not appear sufficient.
\cite{2004MNRAS.350L..52C} pointed out that if the front had a certain finite width, e.g., because of diffusion, the KHI would not develop (though the front is observed to be sharper than the Coulomb collisional m.f.p., so the natural candidate to create such broadening, collisional diffusion, is excluded).
Viscosity, particularly in the anisotropic Braginskii regime, can suppress the growth of KHI by damping small-scale perturbations. For A3667, \citet{2017MNRAS.467.3662I}  found that the observed small-scale smoothness along with the presence of intermediate-scale perturbations of the front shape are consistent with an effective viscosity that is a fraction of the Spitzer value.

\citet{2001ApJ...549L..47V} proposed that the surface tension of a magnetic field aligned with the front --- a field configuration that should naturally arise in the presence of the velocity shear \citep[e.g.,][]{2011ApJ...743...16Z} --- can suppress KHI at small scales and preserve the sharpness of the front within the observed sector. 

In this section, 
we consider a simple model that assumes that the magnetic field along the cold front suppresses the instability, as originally proposed by \citet{2001ApJ...549L..47V}, but assuming a different merger geometry supported by the XRISM velocity measurements.
The condition for suppressing KHI at a shear interface is met when the magnetic tension exceeds the destabilizing ram pressure of the shear flow. This criterion is given by
\begin{equation}
\frac{B^2}{4\pi} \geq \rho \, (\Delta v)^2,
\end{equation}
\color{black}
which can be rearranged to express the minimum magnetic field strength $B_{\rm crit}$ required to stabilize the interface:
\begin{equation}
    B_{\rm crit} = \Delta v \sqrt{4\pi {\rho}}.
\end{equation}
Here, $\Delta v$ represents the velocity shear across the interface.

To estimate $B_{\rm crit}$, we adopt $\Delta v = 535^{+167}_{-154}$~km~s$^{-1}$ as measured between regions~A and B, and use the effective density $\rho_{\rm avg} = {\rho_{\rm in}\,\rho_{\rm out}}/({\rho_{\rm in}+\rho_{\rm out}})$ following equation (14) of \citet{2002AstL...28..495V}. We take electron densities of $n_{\rm in} = 3.2\times10^{-3}$~cm$^{-3}$ and $n_{\rm out} = 8.2\times10^{-4}$~cm$^{-3}$ for the inner and outer regions, respectively \citep{2001ApJ...551..160V}.
Substituting these values into the above equation yields $B_{\rm crit} = 6.8^{+2.1}_{-2.0}~\mu\mathrm{G}$, corresponding to a lower limit of $>$4.8$~\mu\mathrm{G}$ (1$\sigma$ confidence level). 
This value represents a lower limit on the magnetic field strength required to fully suppress KHI (i.e., if magnetic tension is the only stabilizing agent) under our merger geometry assumption, where there is no gas velocity component along the front in the plane of the sky.

We emphasize that this estimate assumes, for consistency with the merger geometry adopted throughout this work, that the velocity contrast arises purely from motions along the line of sight, with no sky–plane component of the shear. Under this simplified assumption, the derived $B_{\rm crit}$ should be interpreted as 
a rough estimate, because any unmodeled geometrical component of the flow could modify the true tangential shear velocity.
\color{black}

Our $B$\/ estimate is consistent with the value derived for the A3667 front by \cite{2001ApJ...549L..47V} under a different assumption about the merger geometry and the gas flow pattern.
Simulations have also shown that magnetic fields of $\sim$10--20~$\mu$G aligned with the front surface are sufficient to prevent the growth of instabilities \citep[e.g., ][]{2011ApJ...743...16Z}.

In A3667, MeerKAT observations revealed a thin, polarized radio layer around the mushroom feature in the NW region of the cluster, interpreted as magnetic draping \citep{2022A&A...659A.146D}, thus providing observational support for shear‐amplified magnetic fields.
\color{black}

Future comparison with MHD simulations matched to the observed profiles will be crucial to constrain the magnetic field configuration and its role in shaping the front.

\section{Conclusion}

We have presented spatially resolved X-ray spectroscopy of the merging galaxy cluster A3667 with \textit{XRISM}/Resolve. The derived LoS velocity and velocity dispersion maps show large variations, providing direct spectroscopic evidence of merger-driven gas dynamics.

Most notably, we detect a LoS velocity change of $535^{+167}_{-154}$~km~s$^{-1}$ across the cold front interface, between the cooler, denser gas and the outer hot gas. 
The velocity changes again by $\sim$400~km~s$^{-1}$ between the low-entropy gas under the cold front and the more central ICM, suggesting that the gas that forms the cold front is moving along the LoS with respect to the rest of the cluster.

These results support an off-center merger scenario for A3667, in which the cold front is formed by low-entropy gas in the core of one of the merging subclusters that has acquired angular momentum and is sloshing in the plane perpendicular to the sky, as proposed by \citet{2024A&A...689A.173O}. 
There is no significant line-of-sight velocity difference in the core perpendicular to the merger axis, and the 300-400~km~s$^{-1}$ cool core LoS velocity is in the range of prediction from merger simulation. 
The enhanced velocity dispersion of $\sim$420~km~s$^{-1}$ in the gas immediately inside the cold front (region~B) likely arises from a combination of turbulence and the line-of-sight projection of the velocity shear within the sloshing core.

From the absence of the KHI at the front surface and using the measured LoS velocity difference across the front, we estimate a minimum magnetic field strength of $B_{\rm crit} = 
7\pm2
~\mu$G 
if magnetic tension alone is responsible for suppressing KHI. These findings highlight the unique capability of nondispersive microcalorimeter spectroscopy to reveal both the dynamical state and the microphysical conditions of the ICM.


\section*{Acknowledgements}
Y.O. would like to take this opportunity to thank the ``Nagoya University Interdisciplinary Frontier Fellowship'' supported by Nagoya University and JST, the establishment of university fellowships towards the creation of science technology innovation, Grant Number JPMJFS2120.
Y.O. was supported by the Sasakawa Scientific Research Grant from The Japan Science Society.
This work was supported by JSPS KAKENHI grant numbers 
JP20H00157 (K.N.), JP22H00158, JP23H04899, JP25H00672 (Y.F.), and JP25K23398 (S.U.).
The material is based upon work supported by NASA under award number 80GSFC21M0002 (F.M.).
S.U. acknowledges support by Program for Forming Japan's Peak Research Universities (J-PEAKS) Grant Number JPJS00420230006. 	
I.Z. acknowledges support from NASA grant 80NSSC18K1684 and partial support from the Alfred P. Sloan Foundation through the Sloan Research Fellowship. Support for J.Z. was provided by the {\it Chandra} X-ray
Observatory Center, which is operated by the Smithsonian Astrophysical Observatory for and on behalf of NASA under
contract NAS8-03060.
This paper employs a list of Chandra datasets, obtained by the Chandra X-ray Observatory, contained in the Chandra Data Collection ~\dataset[DOI: 10.25574/cdc.503]{https://doi.org/10.25574/cdc.503}
\color{black}

\bibliography{apjl}{}
\bibliographystyle{aasjournalv7}


\appendix

\section{Robustness of the SSM analysis in region~A}
\label{sec:ssm_free_fitting}

In the main analysis, we fixed the temperature and abundance of region~A to
$kT = 8.0$~keV and $Z = 0.41$~$Z_{\odot}$, following the results from \textit{XMM-Newton} \citep{2024A&A...689A.173O}. This choice mitigates the degeneracy caused by limited photon statistics and spectral leakage from region~B, as well as possible systematic errors in the telescope PSF model. To assess the robustness of this
choice, we repeated the SSM fitting while leaving the temperature and abundance
in region~A as free parameters.

\begin{figure*}[h]
  
  \centering
  \begin{minipage}{0.40\columnwidth}
    \centering
    \includegraphics[width=\columnwidth]{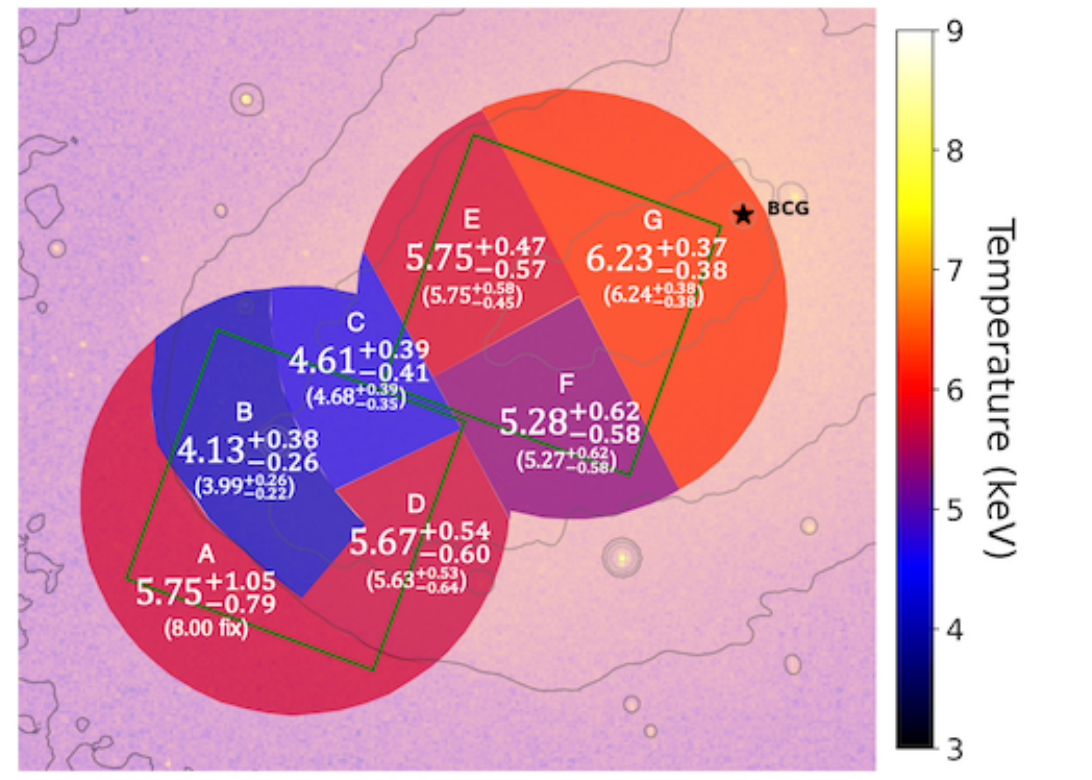}
  \end{minipage}
  \begin{minipage}{0.40\columnwidth}
    \centering
    \includegraphics[width=\columnwidth]{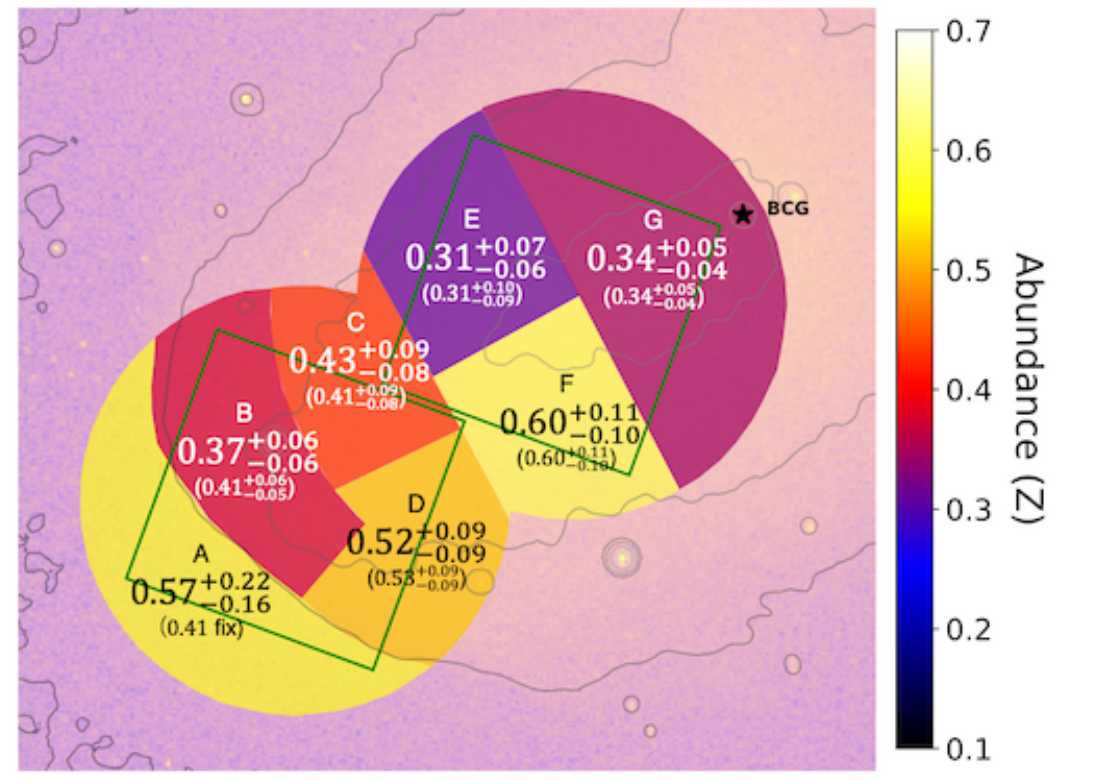}
  \end{minipage}
  
  \centering
  \begin{minipage}{0.40\columnwidth}
    \centering
    \includegraphics[width=\columnwidth]{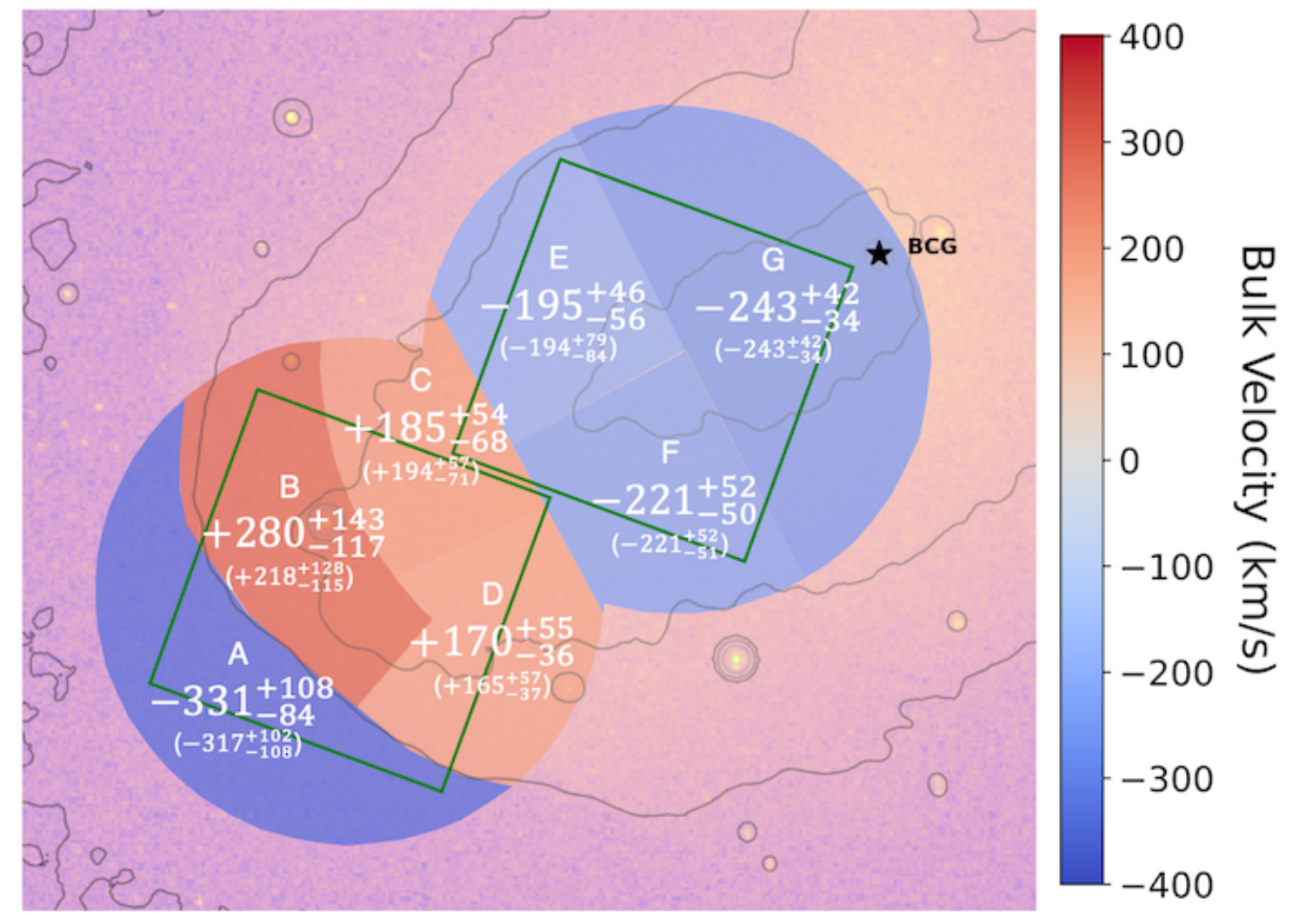}
  \end{minipage}
  \begin{minipage}{0.40\columnwidth}
    \centering
    \includegraphics[width=\columnwidth]{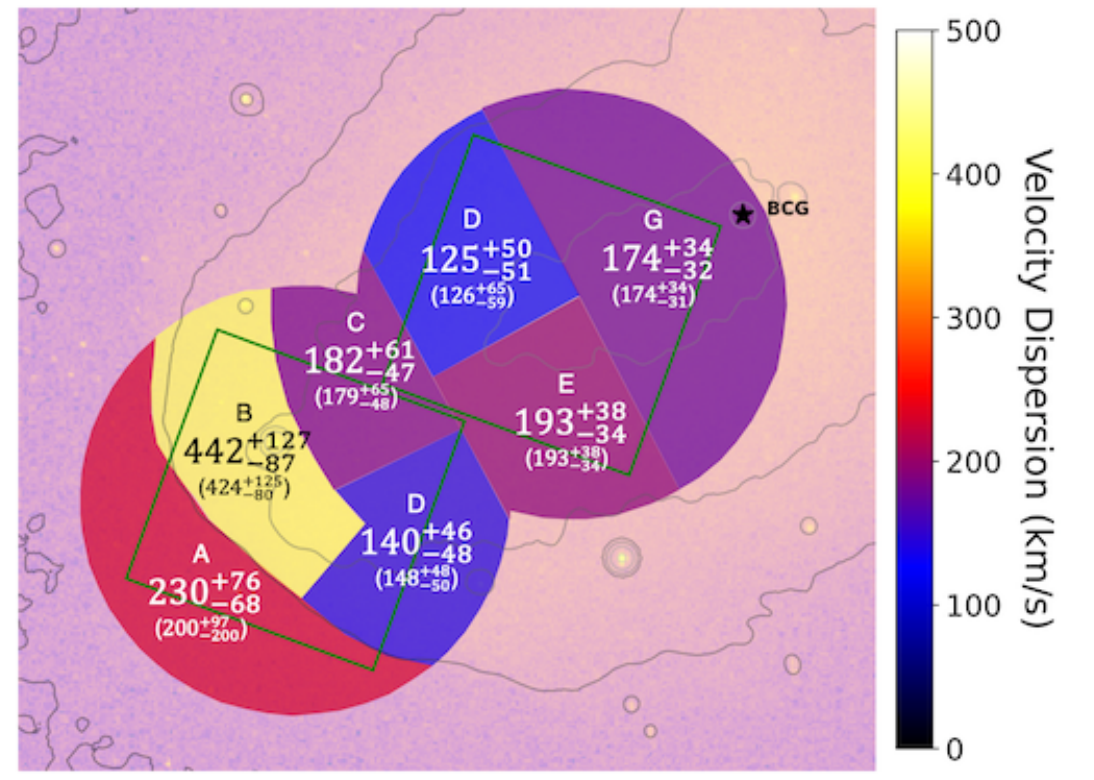}
  \end{minipage}
\caption{Same as Figure~\ref{fig:Resolve_map_ssm}, but with the temperature and abundance parameters in region~A left free. The values shown in parentheses correspond to the results obtained with these parameters fixed
($kT=8.0$~keV and $Z=0.41$~$Z_{\odot}$).}\label{fig:Resolve_map_ssm_free}
\end{figure*}

\begin{figure*}[htpb]
\centering
\includegraphics[width=18cm]{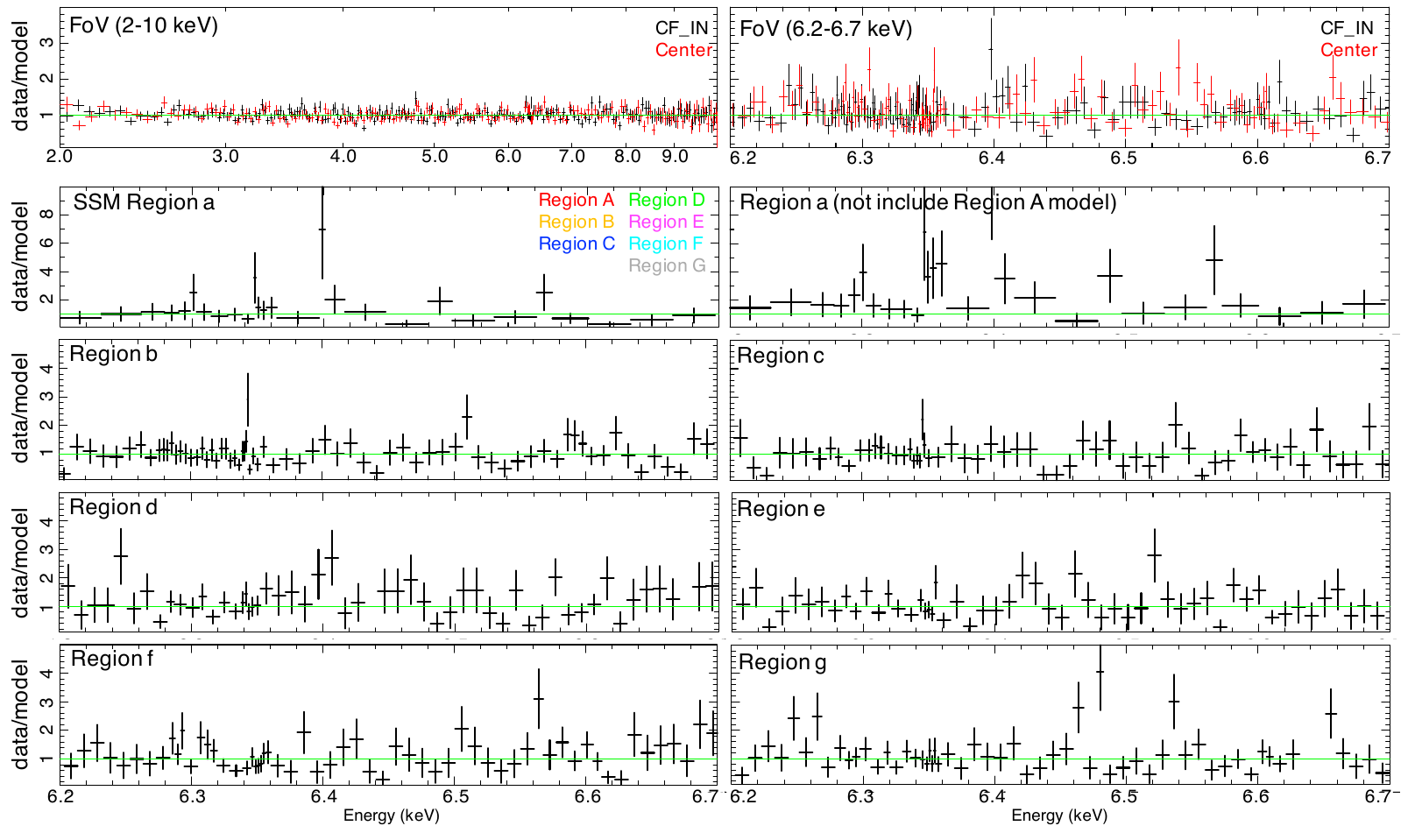}
\caption{Data-to-model ratio plots corresponding to the spectral fits shown in Figure~\ref{fig:Resolve_spectra}. 
\color{black}
{Alt text: Ten line graphs. In the upper left panel, the x axis shows the energy from 2.0 to 10.0 kilo electron volt. The y axis shows the data-to-model ratio of from 0 to 4. In the upper right panel, the x axis shows the energy from 6.2 to 6.7 kilo electron volt. The y axis shows the data-to-model ratio of from 0 to 4.
In the the second row of panels, the x axis shows the energy from 6.2 to 6.7 kilo electron volt. The y axis shows the data-to-model ratio of from 0 to 10.
In the bottom panels, the x axis shows the energy from 6.2 to 6.7 kilo electron volt. The y axis shows the data-to-model ratio of from 0 to 5.
}}\label{fig:Resolve_ratio}
\end{figure*}

Figure~\ref{fig:Resolve_map_ssm_free} shows the resulting maps. In region~A, the free fit yields a temperature of $5.75^{+1.05}_{-0.79}$~keV and an abundance of $0.57^{+0.22}_{-0.16}$~$Z_{\odot}$, indicating a cooler and more metal-rich solution compared to the fixed-parameter analysis. 
As shown in Table~\ref{tab2:Fit parameter in detector regions}, the simple single–temperature fits that ignore SSM mixing yield notably different thermodynamic parameters; for example, region~a is fitted with a temperature of $\sim$4.5~keV, substantially lower than the value obtained from the free-parameter SSM analysis. This demonstrates that PSF and spatial mixing can bias the apparent temperature structure if not modeled properly. 
However, the bulk velocity and velocity dispersion remain consistent within statistical uncertainties, and the overall velocity structure is unchanged. This can be clearly seen in Figure~\ref{fig:Resolve_ratio}, where the spectrum of (detector plane) region~a, excluding the region~A model (i.e., the sum of all other models), exhibits positive residuals on the high-energy side of the He-like Fe line complex. These residuals require the inclusion of a blueshifted component from region~A, demonstrating that the velocity offset is robust. 
\color{black}

The higher abundance obtained in the free fit is likely an artifact of systematic uncertainties in the SSM. Specifically, the leakage fraction from region~B into region~a is likely underestimated; if the true leakage is larger, the equivalent width of lines for region~A will be underestimated in the SSM analysis. In such a case, the fit would favor an even lower turbulent broadening and a stronger blueshifted velocity component. Therefore, the main conclusions regarding the velocity field remain unaffected. Future observations targeting the vicinity of region~A will be essential for reducing the systematic uncertainties in the SSM and obtaining more precise constraints on the magnetic field strength at the cold front.

\color{black}

\label{sec:ssm_free_fitting}
\begin{table*}[h!]
\begin{center}
\caption{Best-fit Parameters of the detector regions in 2.0--10.0~keV energy range. }{%
\resizebox{0.98\textwidth}{!}{%
\begin{tabular}{@{}lcccccccc@{}}
\hline\noalign{\vskip3pt}
\multicolumn{1}{l}{Parameters} & Regions a & b & c & d & e & f & g  \\ [2pt]
\hline\noalign{\vskip3pt}
$kT$~(keV) & $5.08^{+0535}_{-0.46}$ & $4.54^{+0.18}_{-0.15}$ & $4.75^{+0.19}_{-0.20}$ & $5.20^{+0.35}_{-0.33}$ & $5.55^{+0.28}_{-0.35}$ & $5.53^{+0.27}_{-0.27}$ & $6.12^{+0.31}_{-0.30}$  \\
Abundance~($Z_{\rm{\odot}}$) & $0.41^{+0.09}_{-0.08}$ & $0.41^{+0.03}_{-0.03}$ & $0.43^{+0.05}_{-0.04}$ & $0.49^{+0.06}_{-0.06}$ & $0.35^{+0.04}_{-0.04}$ & $0.47^{+0.05}_{-0.04}$ & $0.35^{+0.04}_{-0.04}$  \\
Redshift ($\times$10$^{-2}$) & $5.498^{+0.035}_{-0.032}$ & $5.636^{+0.017}_{-0.016}$ & $5.625^{+0.014}_{-0.013}$ & $5.607^{+0.020}_{-0.012}$ & $5.514^{+0.012}_{-0.010}$ & $5.507^{+0.011}_{-0.010}$ & $5.4934^{+0.0113}_{-0.0086}$ \\
Relative Velocity (km~s$^{-1}$) & $-231^{+104}_{-96}$ & $+184^{+51}_{-48}$ & $+150^{+42}_{-38}$ & $+96^{+61}_{-37}$ & $-183^{+29}_{-37}$ & $-203^{+32}_{-29}$ & $-242^{+34}_{-26}$ \\
Velocity dispersion (km~s$^{-1}$) & $280^{+83}_{-73}$ & $328^{+44}_{-40}$ & $238^{+36}_{-32}$ & $251^{+53}_{-56}$ & $162^{+33}_{-31}$ & $193^{+26}_{-24}$ & $171^{+30}_{-27}$ \\
C-stat/$d.o.f$ & $5456/15994$ & $12882/15994$ & $10119/15994$ & $8893/15994$ & $9785/15994$ & $10110/15994$ & $11220/15994$ \\
[2pt]
\hline\noalign{\vskip3pt}
\end{tabular}
}} 
\label{tab2:Fit parameter in detector regions}
\end{center}
\end{table*}

\end{document}